\documentclass[eqsecnum,aps,showpacs,superscriptaddress,unsortedaddress]{revtex4}

\def\beq{\begin{equation}}
\def\eeq{\end{equation}}

\begin{document}

\title{A New Paradigm in Chaotic Ray Scattering}
\author{G. Castaldi}
\affiliation{Wavesgroup, University of Sannio at Benevento, Italy}
\author{V. Fiumara}
\affiliation{Wavesgroup, D.I.$^{3}$E., University of Salerno, Italy}
\author{V. Galdi}
\affiliation{Wavesgroup, University of Sannio at Benevento, Italy}
\author{V. Pierro}
\affiliation{Wavesgroup, University of Sannio at Benevento, Italy}
\author{I.M. Pinto}
\affiliation{Wavesgroup, University of Sannio at Benevento, Italy}
\date{\today}


\begin{abstract}
We introduce a new paradigm of  (electromagnetic) 2D  ray-chaos, 
featuring both guided and scattered rays 
in a dielectric layer with exponentially tapered refraction index
backed by an undulated conductive surface, and illustrate
its relevant features.
Numerical simulations of the corresponding full-wave solution
indicate that the system complies with Berry's conjecture in the
asymptotic short wavelength limit.
\end{abstract}

\pacs{05.45.Mt, 42.15.Dp, 42.25.Fx}
\maketitle


\section{Introduction}

Ray-chaos (exponential separation of nearby-originating rays,
due to the geometrical features of the reflecting boundaries,
and/or the refractive properties of the medium) is a lively subject,
as witnessed by an increasing number of topical papers.

Well known paradigms of two-dimensional (henceforth 2D) ray-chaos 
include the Sinai \cite{Sinai_bill} and Bunimovic \cite{Buni_bill}
billiards ({\em internal} boundary value problems)
and the $n$-disk ($n \geq 3$) pinball-scatterer
({\em external} boundary value problem)  \cite{pinball}. 
These systems have been extensively studied 
both theoretically  \cite{Theory_Review}, 
in view of their  relevance in the context
of quantum chaology \cite{Quantum_chaology},
and experimentally (see e.g. \cite{Stockmann}, and \cite{Smilanski}), 
thanks to the formal equivalence between 
the 2D Schr\"{o}dinger equation for a point particle in an infinite-wall potential,
and the 2D (scalar) Helmholtz equation 
with perfectly conducting boundaries of Electromagnetics, 
which makes an experimental approach based on microwave measurements viable.

In this paper we introduce and analyze a new intriguing paradigm
in ray-chaotic electromagnetic (henceforth EM) boundary-value problems.
In its simplest (2D) form, the problem's geometry is sketched in fig. 1, 
and consists of a dielectric layer with exponentially tapered refraction index
matched at $z=0$ to the outer vacuum  (viz., $n(0)=1$),
and backed by a smooth  perfectly-conducting 
periodic (undulated) surface $z=\zeta(x)$.   

This geometry bears several elements of novelty. 
It is, perhaps, the simplest (2D) internal/external electromagnetic ray-chaotic boundary value problem 
featuring {\em both} guided  {\em and} scattered rays at the same time. 
It also  includes a (linear, isotropic, time-invariant, non-dispersive)
dielectric layer as a basic ingredient 
at variance of billiards and pinballs, which are pure 
perfect-conductor-and-vacuum problems.

The 2D ray trajectories  bear some similarity 
to the space-time ones  
of the well known (one-dimensional) Fermi-Pustil'nikov billiard \cite{Fermi_bill}, 
In this analogy, the gradient of the refractive-index 
plays the role of gravity, bending the ray-paths downward in the dielectric layer, 
while the undulated surface plays the role of the vibrating table, 
changing the slope of the ray at each bounce.
$$~$$

\section{Ray Tracing}

The system in Fig. 1 can be parametrized 
in terms of the dimensionless variables
$n_f=n(-h)$,  $a/h$, $\Delta/a$.
In this section we illustrate the main relevant results
of straightforward ray-tracing simulations 
run  for the geometry of Fig. 1, with 
$a/h=1$ and $\Delta/a\!=\!2\pi/5$.

In Fig.  2  three   examples are shown
of  (multi-hop) ray-paths originating from rays 
with close-by incidence points ($\Delta x \sim 10^{-3} a$) 
and the same  incidence angle ($\theta_i=\pi/12$).
In all three cases, a rapidly {\em increasing} ray separation
is observed,  resulting into completely different exit-angles.
It will be shown soon that the ray separation increases
exponentially in time, as the rays bounce back and forth
through  the dielectric layer.
 
In Fig. 3  the exit-angle is shown vs. the (scaled) incidence point position
$\bar{x}=x/a$, $a$ being the spatial period of the undulated surface,
for a fixed incidence angle $\theta_i=\pi/12$,
and a few representative values of $n_f$.
Intervals of distinctly {\em irregular} behaviour are observed.
The measure of the set of incidence points 
corresponding to regular behaviour
is seen to depend on the value of $n_f$, 
and shrinks  to zero as $n_f$ is increased.
On the other hand, at  fixed $n_f$, 
the measure of the set of incidence points 
corresponding to regular behaviour
is found  to be essentially  {\em independent}
of the incidence angle, 
as exemplified in Fig. 4, where $n_f=15$
and $\theta_i=k\pi/12$, $k=1,3,5$.

Closer investigation reveals that smaller and smaller 
intervals of regular behaviour are hidden
within {\em arbitrarily small} intervals of  $\bar{x}$,
as shown in Fig. 5 for $\theta_i=\pi/12$.

As anticipated, in the {\em irregular} regime,
the separation  between neighbouring incident rays
increases {\em exponentially}, as the rays bounce 
back and forth through  the dielectric layer.
This is shown in  Fig. 6, where the logarithm of the 
ray-separation (in units of its initial value) 
is displayed as a function
of time (sec) for $n_f=15$, and $\theta_i=k\pi/12$, $k=1,3,5$.
$$~$$

\section{Full Wave Analysis:  Results}

A full-wave (Floquet mode) analysis 
(which is omitted here for brevity, and 
will be found in a forthcoming paper)
suggests that in the limit of increasingly {\em large} $a/\lambda_0$,
$\lambda_0$ being the field wavelength in vacuum
both the internal $( z \leq 0)$ 
and the external ($z >0$) fields 
eventually comply with Berry's conjecture \cite{Berry_conj}.

This is illustrated in Fig.s  7 and 8, which refer to the
special (scalar) case where the electric field has only a nonzero
component orthogonal to the $xz$ plane,
and all other parameters hold the same values as in Sect. 2.

The field scattered into the half-space $z>0$
is indeed a superposition  of (propagating or evanescent) plane waves 
whose wave-vectors span the characteristic (grating-lobe) directions \cite{Beck_Spizz}
\beq
\sin \theta_p = \sin \theta_i - n_f\frac{p\lambda_0}{a},~~~p \in {\cal Z}.
\label{eq:grating_lobes}
\eeq
with (complex) amplitudes $\Gamma_p E_i$
where $E_i$ is the (complex) amplitude of the incident plane wave,
and $\theta_i$ is the incidence angle. 
It is seen from (\ref{eq:grating_lobes})
that the number of propagating scattered plane waves is $\propto n_fa/\lambda_0$.
Numerical simulations  further show 
that upon increasing $a/\lambda_0$, the distribution of the phase 
among the pertinent  (complex) coefficients $\Gamma_p$
becomes more and more evenly distributed in $(0,2\pi)$.

On the other hand, as seen from Fig. 7, 
the agreement between the (averaged \cite{average}, normalized)
spatial autocorrelation function
$\langle E[\vec{r}+\hat{x}(s/2)]E^*[\vec{r}-\hat{x}(s/2)] \rangle$ 
of the internal ($z \leq 0$) field
and the $J_0(n(z)k_0|\vec{s}|)$ Bessel function limit-form 
predicted by  Berry's conjecture 
is remarkably good already at $a/\lambda_0=6$.

It is also noted that the complex coefficients $\Gamma_p$
show a {\em strong} sensitivity to the incidence angle $\theta_i$,
as illustrated in Fig. 8 for  $a/\lambda_0=9$.
In Fig. 8  the case of a plain sinusoidal surface is also shown
($n_f=1$), for which the $\Gamma_p$s
are seen to depend fairly {\em smoothly} on the incidence angle.
This peculiar behaviour 
has a special notable consequence,
also suggested by numerical simulations.
In the limit where $a/\lambda_0 \rightarrow 0$
the spread of the distribution of field values at any given point 
corresponding to a large number of different incidence angles 
in a sector of  aperture $\Delta\theta_i$.
also remains ${\cal O}(\Delta\theta_i)$,
and the mean  changes with the chosen field point.
On the other hand,
in the opposite limit  where $a/\lambda_0 \rightarrow \infty$,
for {\em any} nonzero $\Delta\theta_i$, albeit small,
the spread of the distribution {\em blows-up} to a finite value,
while both the spread and the mean of the distribution
become almost independent from the chosen field point.
This scenario is likely to occur in all wave-scattering problems 
exhibiting chaos in the $\lambda \rightarrow 0$ (ray) limit.
$$~$$ 
\section{Conclusions - Application Potential}

Far from being a mere exercise of pure academical interest,
the proposed scattering system  holds the potential for several applications
of practical interest.

The findings  presented above suggest
that an incident  narrow coherent beam of electromagnetic radiation
would be scattered off the system {\em incoherently} and {\em isotropically}
if $a/\lambda_0 \gg 1$. 
Thus  the proposed  system might be used as a {\em coating} 
for radar-puzzling applications.

The multi-hop nature of the ray-paths also suggest that 
upon addition of suitably {\em small} \cite{lossy} dielectric and/or backing-surface losses
the system might act as a smart radio-frequency absorber,
featuring both fairly {\em uniform} power deposition in the dielectric bulk,
and nicely small power reflection-coefficient 
for a {\em wide} interval of incidence angles.
$$~$$

\section*{Acknowledgements}

We acknowledge several stimulating discussions with professor Leo B. Felsen 
(Boston University, Boston MA, USA) and professor Giorgio Franceschetti
(University of Naples, IT, and UCLA, Los Angeles CA, USA).
This paper is dedicated to the memory of dr. Maurizio Mangrella.





\newpage
\begin{center}
{\bf Captions to the Figures}
\end{center}
$$~$$
Fig 1 - Sketch of problem's geometry. A perfectly conducting undulated surface
with peak-to-peak height $\Delta$ and spatial period $a$ is topped by a dielectric
layer with maximum thickness $h$. The refractive index of the layer matches
that of the vacuum at $z=0$ and grows exponentially with depth up to a value
$n_f$ at $z=-h$.
$$~$$
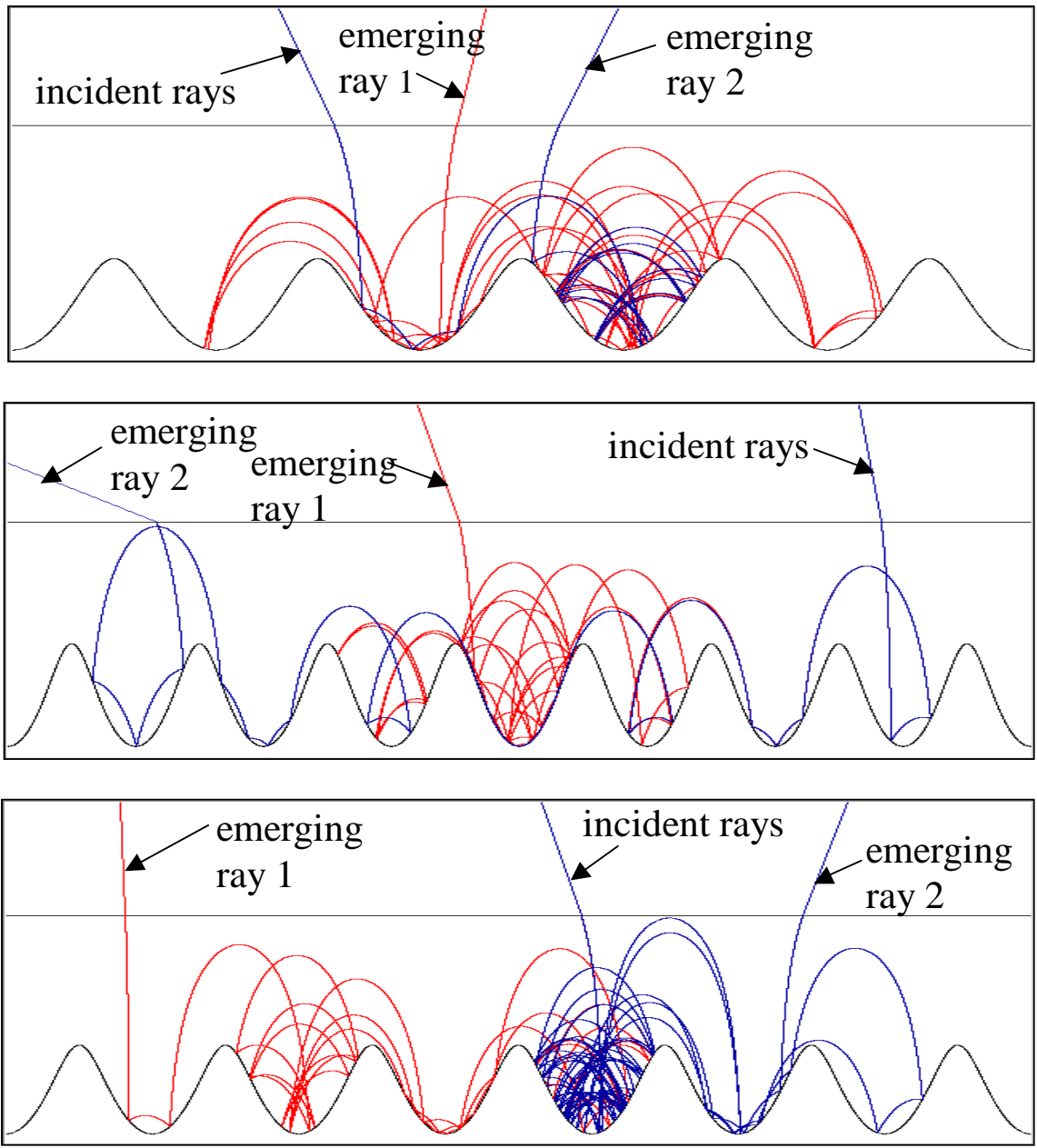
Fig. 2 - Multi-hop ray-paths originating from close-by incidence points
display rapidly increasing separations, and eventually emerge with
widely different exit angles.
$$~$$
Fig. 3 - Exit angle (deg.) vs. scaled incidence point position $\bar{x}=x/a$
for fixed incidence angle ($\theta_i=\pi/12$) and different values of $n_f$:
$n_f=1.4~\mbox{(top)}$, $n_f=30~\mbox{(mid)}$, $n_f=90~\mbox{(bottom)}$.
$$~$$
Fig. 4 - Exit angle (deg.) vs. scaled incidence point position $\bar{x}=x/a$
for fixed $n_f$ and different incidence angles. 
$\theta_i=\pi/12~\mbox{(top)}$, $\theta_i=3\pi/12~\mbox{(mid)}$, $\theta_i=5\pi/12~\mbox{(bottom)}$. 
$$~$$
Fig. 5 - Exit angle (deg.) vs. scaled incidence point position $\bar{x}=x/a$ for
$n_f=15$ and $\theta_i=\pi/12$.  Intervals corresponding to regular behaviour 
are observed at smaller and smaller scales (top to bottom).
$$~$$
Fig. 6 - Natural logarithm of separation between close-by incident rays
as a function of time for $n_f=15$ and different incidence angles:
$\theta_i=\pi/12~\mbox{(left)}$, $\theta_i=3\pi/12~\mbox{(mid)}$, $\theta_i=5\pi/12~\mbox{(right)}$.
$$~$$
Fig. 7 - Full wave analysis with $a/\lambda_0=6$. 
Averaged normalized  spatial autocorrelation function of the internal $(z \leq 0)$ field.
$$~$$
Fig. 8 - Full wave analysis with $a/\lambda_0=9$. 
Top: modulus (left) and phase (right) of  $\Gamma_p$ 
for two close-by incidence angles. 
Bottom:  the same for  a plain ($n_f=1$)  undulated surface.


\begin{thebibliography}{99}
\bibitem{Sinai_bill}{Ya. G. Sinai,  Russian Math. Surveys, {\bf 25}, 137 (1970).}
\bibitem{Buni_bill}{L.A. Bunimovich, Chaos, {\bf 1}, 187 (1991).}
\bibitem{pinball}{C. Bercovich, U. Smilanski and G.P. Farmelo, Eur. J. Phys., {\bf 12}, 122 (1991).}
\bibitem{Theory_Review}{H.J Stockmann, {\em Quantum Chaos: an Introduction}, Cambridge University Press, Cambridge UK (1999).}
\bibitem{Quantum_chaology}{M.V. Berry,  Proc. Roy Soc. London, {\bf A413}, 183 (1987).}
\bibitem{Stockmann}{G. Veble, U. Kuhl, M. Robnik, H.J. Stöckmann, J. Liu and M. Barth,
Prog. Theor. Phys. Suppl. {\bf 139}, 283 (2000).}
\bibitem{Smilanski}{T. Kottos, U. Smilansky, J. Fortuny and G. Nesti, Radio Science, {\bf 34} 747, (1999).}
\bibitem{Fermi_bill}{See, e.g., A.J. Lichtenberg and M.A. Liebermann, {\em Regular and Stochastic Motion}, Springer Verlag, NY (1983).}
\bibitem{Berry_conj}{M.V. Berry,  J. Phys. {\bf A10}, 2083 (1977).}
\bibitem{average}{The average is taken over several (local)  wavelenghts - see, e.g., \cite{Berry_conj}.}
\bibitem{Beck_Spizz}{P. Beckmann and A. Spizzichino, {\em The Scattering of Electromagnetic Waves from Rough Surfaces}, Pergamon Press, NY (1963), ch. 4.}
\bibitem{lossy}{The best performance is obtained when  the losses are small enough to allow
the existence of sufficiently long multi-hop ray-paths.}
\end{thebibliography}
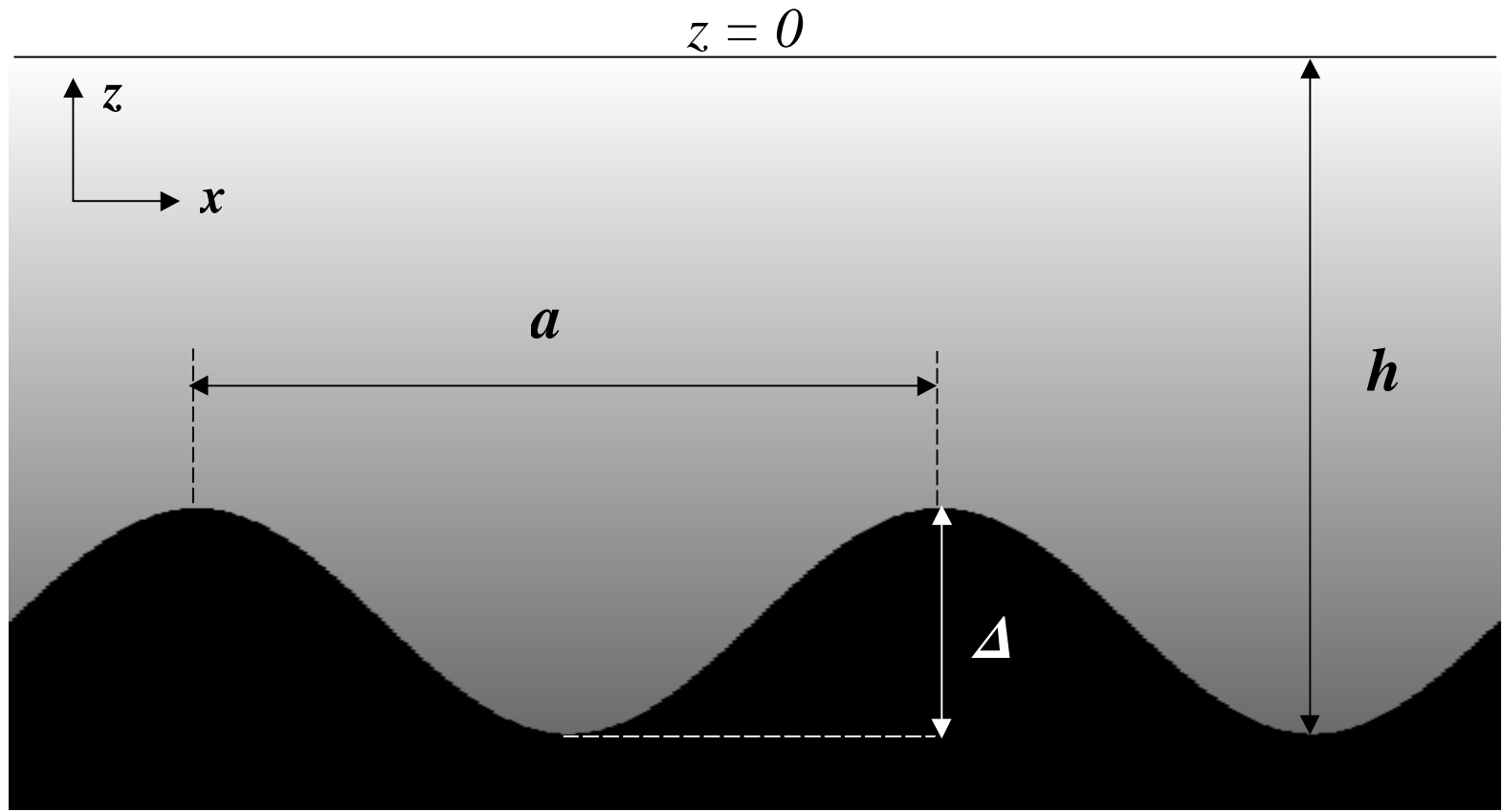
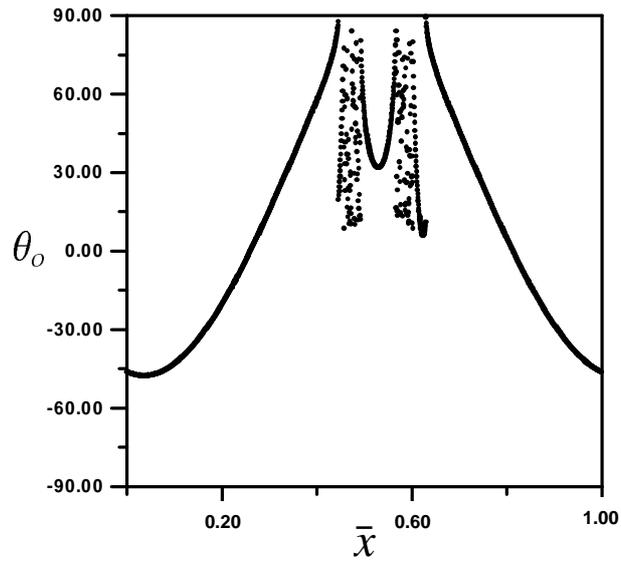
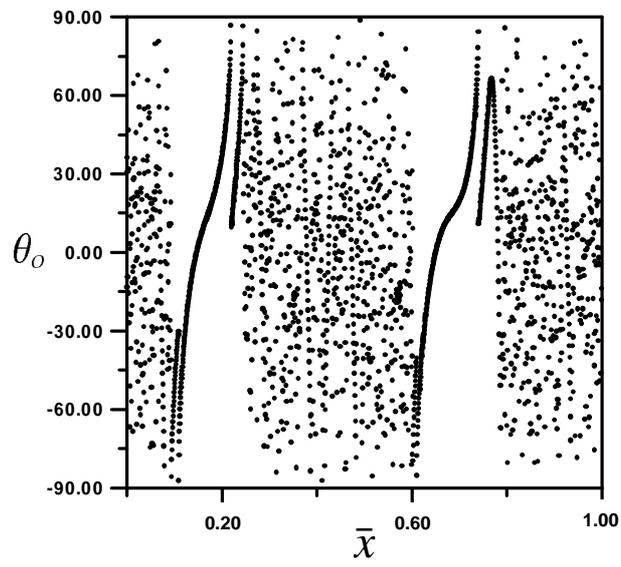
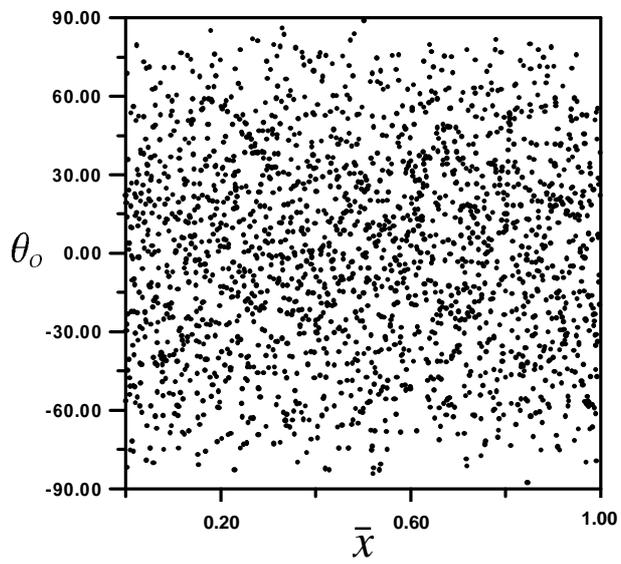
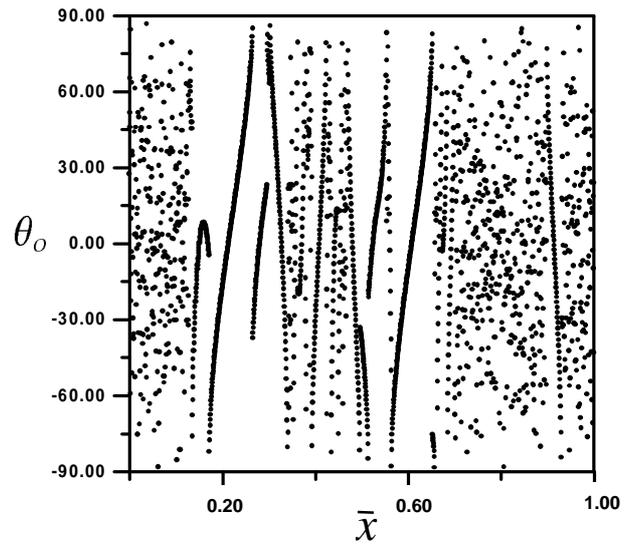
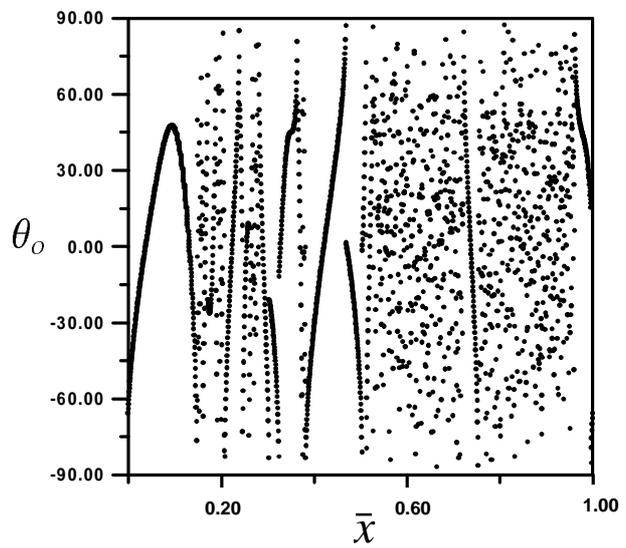
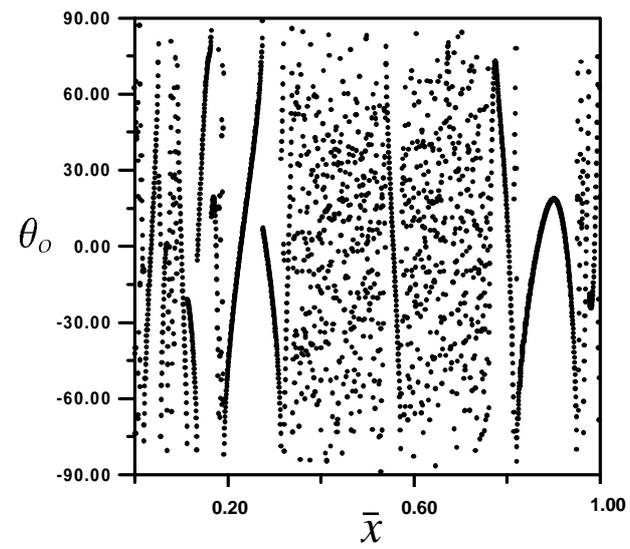
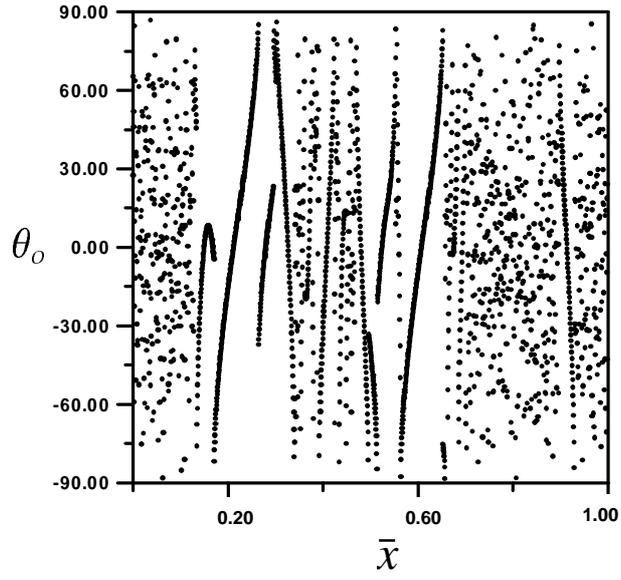
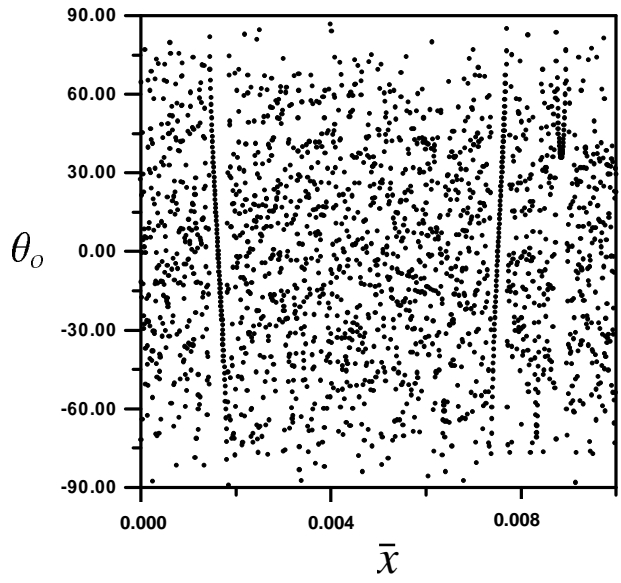
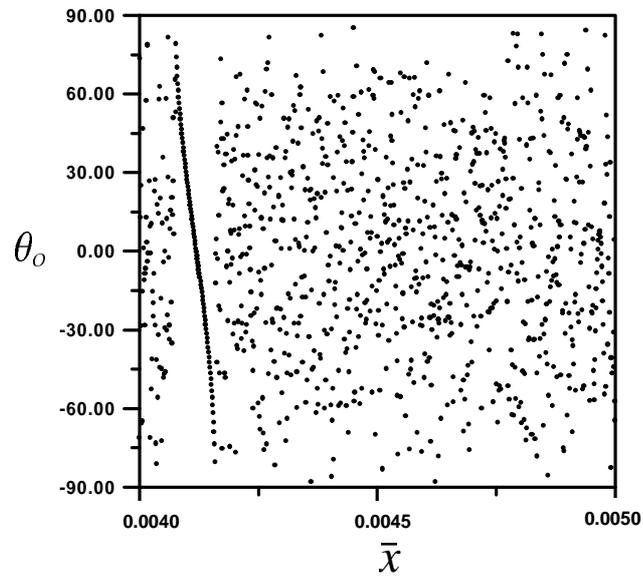
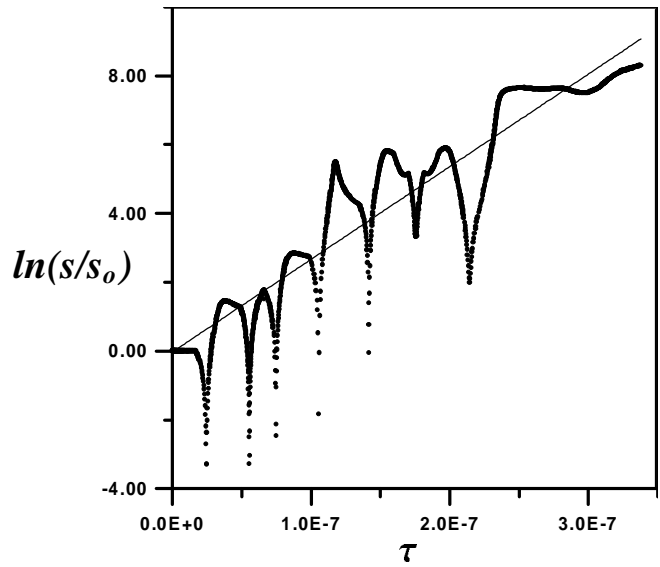
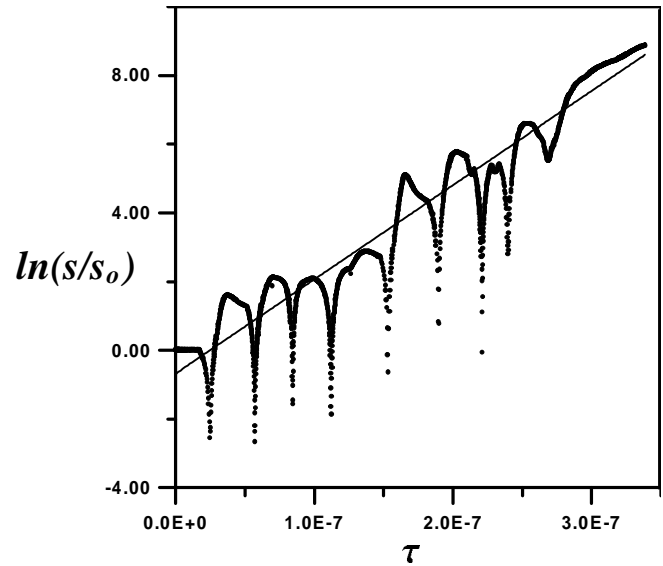
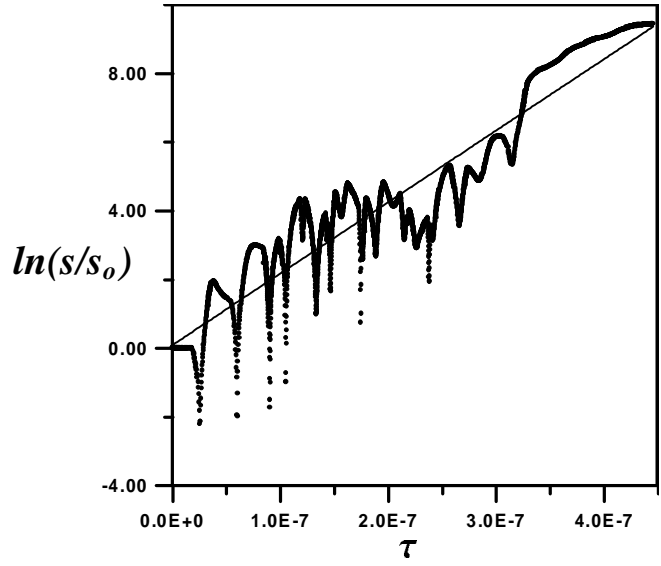
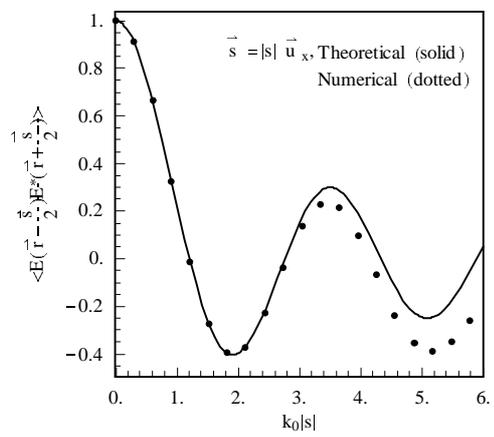
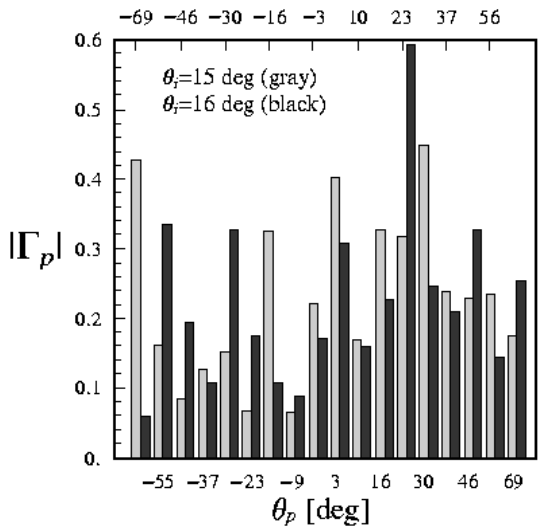
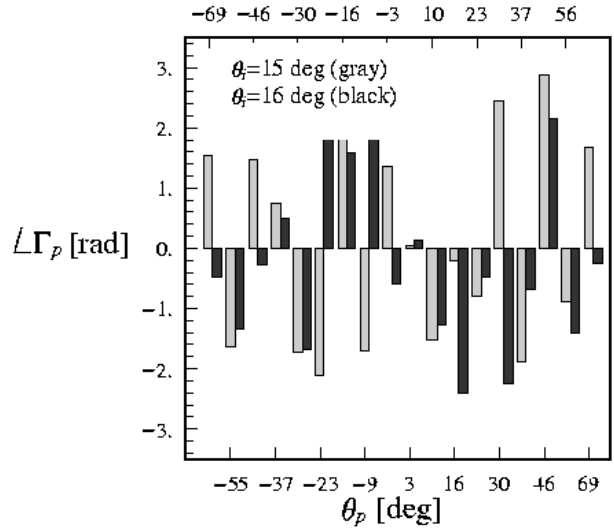
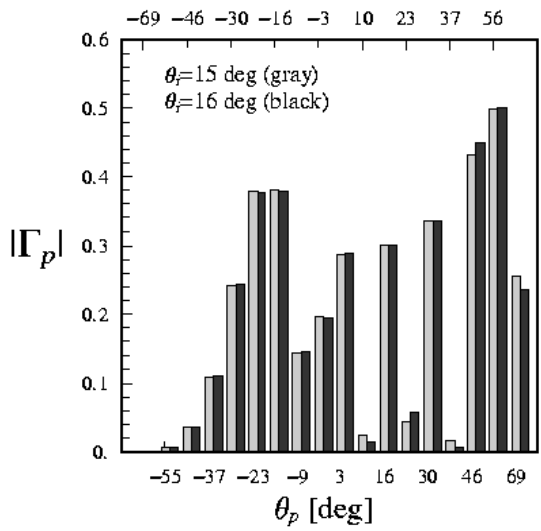
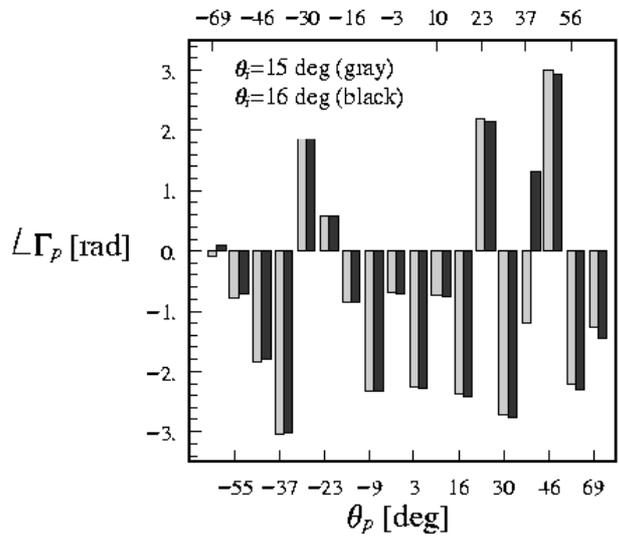
\end{document}